\documentclass[aps,prd,groupedaddress,showpacs,graphicx,nofootinbib]{revtex4}
\usepackage{amssymb,graphics,graphicx,color,amsmath}

\begin{document}

\title{Inflaton as the Affleck-Dine Baryogenesis Field in Hilltop Supernatural Inflation}

\author{Chia-Min Lin$^{1}$}\email{cmlin@ncut.edu.tw}
\author{Kazunori Kohri$^{2}$}\email{kohri@post.kek.jp}

\affiliation{$^1$Fundamental Education Center, National Chin-Yi University of Technology, Taichung 41170, Taiwan}
\affiliation{$^2$Theory Center, IPNS, KEK, Tsukuba 305-0801, Japan}
\affiliation{$^2$The Graduate University of Advanced Studies (Sokendai), Tsukuba 305-0801, Japan}
\affiliation{$^2$Kavli IPMU (WPI), UTIAS, The University of Tokyo, Kashiwa, Chiba 277-8583, Japan}

\date{Draft \today}

\begin{abstract}
In this paper, we investigate the parameter space in the framework of hilltop supernatural inflation in which the inflaton field can play the role of Affleck-Dine field to produce successful baryogenesis. The suitable value of reheating temperature could coincide with the reheating temperature required to produce lightest supersymmetric particle dark matter. The baryon isocurvature perturbation is shown to be negligible. We consider $p=3$, $p=4$, and $p=6$ type III hilltop inflation and discuss how to connect the models to supersymmetric theories. 

\end{abstract}
\maketitle

\large
\baselineskip 18pt
\section{Introduction}

Recently, Planck has given a strong support in favor of inflation \cite{Starobinsky:1980te, Sato:1980yn, Guth:1980zm} (see e.g. \cite{Lyth:2009zz} for a textbook review), the stage of quasi-de Sitter expansion of the Universe prior to the conventional hot stage. While the concrete inflationary scenario realized in nature has not been identified yet, most models of inflation deal with one or many scalar field(s) dubbed inflaton(s). During inflation, the inflaton slowly rolls down the slope of its potential. This potential is required to be substantially flat, or, otherwise speaking, the slow-roll parameters must be small. As inflation ends, the inflaton energy density gets transferred to the Standard Model species.

Inflation can be naturally realized in supersymmetry (SUSY)-thanks to the presence of flat directions in SUSY, along with which inflation can take place. In the present paper, we consider a supersymmetric realization of hybrid inflation called supernatural inflation. However, in its original form, the latter leads to the blue tilted spectrum of primordial scalar perturbations-in conflict with the Planck data. This motivates to promote the original scenario to the hilltop supernatural inflation, which can avoid the conflict with cosmological observations. The hilltop supernatural inflation is the focus of the present work.

SUSY also proved to be advantageous for building the models, which accommodate a dark matter candidate and/or lead to the baryon asymmetry of the Universe. In the context of SUSY, baryon asymmetry is most commonly generated by the Affleck-Dine mechanism \cite{Affleck:1984fy, Dine:1995kz}. The latter typically requires the existence of flat directions in the field space, which is a natural consequence of SUSY. The dark matter candidate is realized in the form of the lightest supersymmetric particle (LSP).

The interplay between inflation and Affleck-Dine baryogenesis has been covered in the literature, for example \cite{Kamada:2010yz,BasteroGil:2011cx,Bastero-Gil:2014oga,Li:2014jqa,Kawasaki:2015cla,Yamada:2015xyr}. An interesting possibility is to investigate whether the inflaton may play the role of the Affleck-Dine field \cite{Charng:2008ke, Hertzberg:2013jba, Hertzberg:2013mba, Unwin:2015wya, Evans:2015mta, Cline:2019fxx}. In the present work, we develop this idea and apply it to the hilltop supernatural inflation. In this unified framework, we also discuss the production of LSP, which is assumed to play the role of dark matter.

\section{Hilltop Supernatural Inflation}
\label{sec1}

The potential for conventional hybrid inflation is given by\footnote{Here the subscripts of $\psi_r$ and $\phi_r$ denote that they are real fields. In the following we are going to consider complex fields.}
\begin{equation}
V=\frac{1}{2}m_{\psi_r}^2 \psi_r^2 +g^2 \psi_r^2 \phi_r^2 + \kappa^2 (\phi_r^2-\Lambda^2)^2,
\label{eq1}
\end{equation}
where $\psi_r$ is the inflaton field and $\phi_r$ is the waterfall field. The effective mass of the waterfall field is
\begin{equation}
m^2_{\phi_r} \equiv V^{\prime\prime}(\phi_r)=2g^2 \psi_r^2 -4\kappa^2 \Lambda^2.
\label{eq2}
\end{equation}
During inflation, the field value of $\psi_r$ gives a large positive mass to $\phi_r$, therefore, it is trapped to $\phi_r=0$ and the potential 
during inflation is of the form 
\begin{equation}
V=V_0+\frac{1}{2}m_{\psi_r}^2 \psi_r^2,
\label{eq3}
\end{equation}
where $V_0=\kappa^2 \Lambda^4$.
The end of inflation is determined by $m^2_{\phi}=0$ when the waterfall field starts to become tachyonic which implies
\begin{equation}
\psi_{r,end}=\frac{\sqrt{2}\Lambda \kappa}{g}
\label{eq4}
\end{equation}

Supernatural inflation is basically hybrid inflation in the framework of SUSY \cite{Randall:1995dj}. During inflation, the potential of the inflaton $\psi$ with mass $m$ is given by
\begin{equation}
V=V_0+\frac{1}{2}m^2 \psi^2
\end{equation} 
which has the same form as Eq.~(\ref{eq3}).
The advantage of supernatural inflation is that the tuning of model parameters is automatic due to SUSY and the model provides more connections to particle physics. The idea is to consider TeV scale SUSY breaking which can be realized by $V_0=M_S^4$ where $M_S \sim 10^{11}\mbox{GeV}$ is the gravity mediated SUSY breaking scale and $m \sim \mbox{TeV}$ is the soft mass. 

By assuming $V \sim V_0$, we can obtain 
\begin{equation}
\psi =\psi_{end}e^{\frac{Nm^2M_P^2}{V_0}} \equiv \psi_{end}e^{N \eta_0}
\label{sne}
\end{equation}
and
\begin{equation}
\left( \frac{\psi}{M_P} \right)^2=\frac{1}{12\pi^2 P_R}\left( \frac{V_0}{M_P^4} \right)\frac{1}{\eta_0^2}.
\label{snp}
\end{equation}
The spectral index is given by
\begin{equation}
n_s=1+2\eta_0.
\label{snn}
\end{equation}
Since $\eta_0>0$, one results with the blue spectral index of primordial scalar perturbations, i.e., $n_s>1$ which has been ruled out by the latest Planck 2018 data \cite{Akrami:2018odb}. This motivates us to modify the model and introduce hilltop supernatural inflation \cite{Lin:2009yt, Kohri:2013gva, Lin:2010zzk, Kohri:2010sj}. In this model, the spectral index can be of the observed value. It can also evade both thermal and nonthermal gravitino problem \cite{Kohri:2010sj}, produce primordial black holes \cite{Alabidi:2009bk}, or induce gravity waves \cite{Alabidi:2012ex}. In the following sections, we will also show that by modifying supernatural inflation into hilltop supernatural inflation, the inflaton field value at the end of inflation becomes larger [compare Eq.~(\ref{a7}) with Eq.~(\ref{snp})]. This difference is crucial when we discuss baryogenesis.

We now consider a term in the superpotential which is responsible to reduce the spectral index, 
\begin{equation}
W= a \frac{\psi_1 \psi_2 \cdots \psi_p}{M_P^{p-3}},
\end{equation}
for $p$ different superfields $\psi_p$,
and $D$-flat direction for $\psi_i$ $(|\psi_1|=|\psi_2| = \cdots =|\psi_p|)$.
Here $a$ is a coupling constant which could be complex with $|a| \sim \mathcal{O}(1)$.
The real inflaton field is a linear combination of the scalar component of the $p$ fields,
\begin{equation}
\psi = \sqrt2 (\psi_1 + \cdots + \psi_p)/\sqrt{p}, \qquad
\psi_i = \psi/\sqrt{2p}.
\end{equation}
The scalar potential during inflation in terms of the kinetic normalized field $\psi$ 
is given as
\begin{equation}
V = V_0 + \frac12 m_\psi^2 \psi^2 - \left(a A \frac{\psi^p}{(2p)^\frac{p}2 M_P^{p-3}} + c.c\right) +
 \frac{p|a|^2 \psi^{2(p-1)}}{(2p)^p M_P^{2(p-3)}},
\end{equation}
where $m_\psi^2 = \sum m_{\psi_i}^2/p$ for the average of the SUSY breaking squared masses of $\psi_i$,
and $A$ is a SUSY breaking parameter. The SUSY breaking mass $m_\psi$ and $A$ are typical of the order of the gravitino mass $m_{3/2}$ \cite{Enqvist:2003gh, Dine:1995kz}.
The detailed values of these parameters depend on specific mechanisms of SUSY breaking.

We will neglect the last term on the right-hand side and approximate the potential as \cite{Kohri:2007gq}
\begin{equation}
V(\psi)= V_0 \left( 1+\frac{1}{2}\eta_0 \frac{\psi^2}{M_P^2}\right)-\lambda \frac{\psi^p}{M_P^{p-4}},
\label{Vpsi}
\end{equation}
%
where
\begin{equation}
\eta_0 \equiv \frac{m_\psi^2 M_P^2}{V_0}, \qquad
\lambda \equiv \frac{2 a A}{(2p)^\frac{p}2 M_P}  .
\label{lambda}
\end{equation}
We assume that $aA$ is real and positive in the convention that all the $\psi_i$ configuration is aligned to be real.
This model can be analyzed by slow-roll approximation and some relevant results are collected in the Appendix.

\section{Baryogenesis}

After inflation, the inflaton field starts to oscillate and eventually decays when the magnitude of its decay width reaches the Hubble parameter.
An interesting question is whether the inflaton can play the role of AD field \cite{Affleck:1984fy, Dine:1995kz}. The possibility was investigated for supernatural inflation in \cite{Randall:1995dj} but it was shown that it does not work for the inflaton field in that model. A quadratic potential of the inflaton field for chaotic inflation which plays the role of an AD field after inflation is investigated in \cite{Charng:2008ke, Hertzberg:2013jba, Hertzberg:2013mba, Unwin:2015wya}. In \cite{Evans:2015mta}, a similar idea is realized in a large field inflation model where the inflaton is a linear combination of right-handed sneutrino fields is considered. Recently, \cite{Cline:2019fxx} considered the inflaton field as the Affleck-Dine field by adding a nonminimal coupling to gravity in order to evade the severe experimental constraints from tensor to scalar ratio. We investigate the case for hilltop supernatural inflation in the framework of small field inflation in this paper.

Our inflaton field is assumed to be a flat direction; therefore, it may play the role of Affleck-Dine field and produce baryon asymmetry if it carries nonzero $B-L$\footnote{It is the relevant quantum number for a nonvanishing $B$ after sphaleron.}. We consider the cases $p=3$, $p=4$ and $p=6$.\footnote{If R-parity is conserved, there is no flat direction with nonzero $B-L$ in the case $p=5$. See \cite{Gherghetta:1995dv} for a review of flat directions.}   
If the inflaton carries nonzero baryon (or lepton) number, baryon number density is produced when the inflaton starts to oscillate. The oscillation in general is accompanied by a spiral motion due to the potential for the angular direction.  
The baryon number density $n_B$ is given by the angular motion of $\psi$ as
\begin{equation}
n_B=iq(\psi \dot{\psi}^\ast -\dot{\psi}\psi^\ast)=q|\psi|^2\dot{\theta},
\end{equation}
where $q$ is the baryon number carried by the AD field, and we have defined $\psi=|\psi|e^{i \theta}$.
The evolution of the baryon number density is given by
\begin{equation}
\dot{n}_B+3Hn_B=2q\mbox{Im}\left[ \psi \frac{\partial V_A(\psi)}{\partial \psi} \right],
\end{equation}
where $V_A = a A \prod_{i=1}^{p} \psi_i/M_P^{p-3}$.
In order to solve the above equation, we multiply both sides of the equality by the cube of the scale factor $R(t)^3$ and integrating with respect to the time $t$ to obtain
\begin{eqnarray}
n_B &=& \frac{2q}{R(t)^3}\int^{t_{\rm sp}}_{t_{\rm end}}R(t)^3 \mbox{Im}\left[ \psi \frac{\partial V_A(\psi)}{\partial \psi} \right]dt  \\
       &=&  \sum \frac{2q_i}{R(t)^3}\int^{t_{\rm sp}}_{t_{\rm end}}R(t)^3 \frac{aA}{M_P^{p-3}} \mbox{Im}\left[ \psi^p \right]\frac{1}{(2p)^{p/2}}dt,
\end{eqnarray} 
where $q_i$ is a charge of $\psi_i$.
The integration is done in a short period of time from the end of inflation to the onset of the spiral motion of the AD field. Because first, we have assumed that baryon number has been diluted during inflation. Second, the contribution to the integral is small after the onset of spiral motion, since the sign of $\mbox{Im}[\psi^p]$ changes rapidly and the amplitude of $|\psi|$ would shrink due to its decay into other particles (with the baryon number conserved). Just after the end of inflation, the universe is matter dominated and the scale factor goes like $R(t) \propto t^{2/3}$. Unlike the common case where the amplitude of the AD field $|\psi|^p$ decreases with time as $H^{p/(p-2)} \propto r^{-p/(p-2)}$ due to its trapping by a large negative Hubble induced mass, our AD field is the inflaton field.   Therefore its energy density is dominated by the oscillation of the quadratic potential which behaves like matter $|\psi|^2 \propto R(t)^{-3} \propto t^{-2}$; thus, $|\psi^p| \propto t^{-p}$. Hence the integrand is proportional to $t^{(2-p)}$. By using $\psi=|\psi|e^{i \theta}$, the integration gives\footnote{Here the result is obtained for $p \neq 3$. When $p=3$, a factor $\propto \ln \left( \frac{M_P}{\phi_{sp}} \right)$ is obtained which is not very sensitive to the value of $\phi_{sp}$.}
\begin{equation}
n_B=\sum 2q_iaA\left( \frac{|\psi_{\rm sp}^p|}{M^{p-3}} \right) \sin[p\theta_{sp}+arg(A)]\times \frac{1}{3-p}t_{sp}\frac{1}{(2p)^{p/2}}.
\label{baryon}
\end{equation}
Note that if $\sin[p\theta_{sp}+arg(A)]=0$, there will be no baryon number generated. Here we will assume $\sin[p\theta_{sp}+arg(A)] \sim 1$ and $\psi_{\rm sp} \sim \psi_{\rm end}$. In the matter dominated universe, we have $R(t) \propto t^{2/3}$ and $H=\dot{R(t)}/R(t) \sim \frac{2}{3}t^{-1}$. Therefore we obtain
\begin{equation}
n_B=\sum 2q_iaA\left( \frac{|\psi_{\rm end}^p|}{M^{p-3}} \right) \times \frac{2}{3(3-p)H}\frac{1}{(2p)^{p/2}},
\end{equation} 
where the Hubble parameter is evaluated at the end of inflation when the inflaton (as the AD field) starts to oscillate. 
By using the definition of $\lambda$ from Eq.~(\ref{lambda}), the above equation can be written as
\begin{equation}
n_B = \sum q_i \lambda \frac{\psi_{\rm end}^p}{H M_P^{p-4}} \times \frac{2}{3(3-p)},
\end{equation}
In general, it can be expressed as
\begin{equation}
n_B \sim q_A \lambda \frac{\psi_{\rm end}^p}{H M_P^{p-4}},
\end{equation}
when the inflaton (AD field) starts to oscillate. Here $q_A$ is the charge carried by $V_A$. For simplicity, in the following, we will set $q_A \sim 1$. In our model, the mechanism of inflation determines the initial conditions of the AD field and there is no need to assume it in some \textit{ad hoc} way. 

The A-term, which induces the first kick for the rotation,  breaks the CP conservation, and the direction of the rotation  breaks the baryon number. Thus, an initial condition of the initial kick breaks both of them in a Hubble patch. Since the initial kick, the produced baryon number has been conserved. The baryon number density will then be diluted due to the expansion of the universe. The universe will then expand until reheating through the baryon-conserving decay of the inflaton. When reheating happens, $\Gamma \sim H$ and hence the reheating temperature is given by $T_R \sim \sqrt{\Gamma M_P}$. 
The strength of baryon-conserving interactions of the inflaton field is connected to the reheating temperature through the decay width $\Gamma$.
At reheating, the baryon number density is given by
\begin{eqnarray}
n_B &\sim& \frac{\lambda}{H}\frac{\psi_{end}^p}{M_P^{p-4}} \times \left( \frac{\Gamma}{H} \right)^2  \\
       &=& \frac{\lambda}{H} \frac{\psi_{end}^p}{M_P^{p-4}}\frac{\pi^2 g_\ast T_R^4}{90 M_P^2H^2} \label{baryon}
\end{eqnarray}
The baryon asymmetry $Y =n_B/s= n_B/(2\pi^2 g_\ast T_R^3/45)=0.9 \times 10^{-10}$ as required by BBN and cosmic microwave backgound (CMB) gives
\begin{equation}
\frac{n_B}{T_R^3} = \frac{\lambda}{H}\frac{\psi_{end}^p}{M_P^{p-2} H^2}T_R = 1.2 \times 10^{-10}.
\end{equation}
%
%
The reheating temperature is given by 
\begin{equation}
T_R=1.2 \times 10^{-10}\times \frac{H^3M_P^{p-2}}{\lambda \psi_{end}^p}.
\label{reheating}
\end{equation}

We assume the SUSY breaking scale such that $m_\psi$ and $A$ are between TeV and $100$ TeV. We will see that CMB constraints require $\eta_0 \equiv m^2_\psi/V_0 \sim 0.01$ or so; therefore, $10^{-26} \lesssim \frac{V_0}{M_P^4} \lesssim 10^{-24} $. As we will see in the following section, in general higher SUSY breaking scale makes baryogenesis easier, but we do not wish to deviate from TeV scale too much. This is the reason behind choosing this range of energy scales.

In the analysis so far, we were assuming that the AD condensate evolves homogeneously after it is formed. In general, there is a possibility that the AD condensate becomes unstable with respect to spacial perturbations and turns into nontopological solitons called Q-balls \cite{Coleman:1985ki, Kusenko:1997zq, Kusenko:1997ad, Enqvist:1997si, Enqvist:1998en}. If Q-balls are formed, our scenario for the evolution of the universe may need to be modified. Q-balls are not formed if $m_{\psi} \gg m_{1/2}$, where $m_{1/2}$ is the mass scale for the gauginos \cite{Allahverdi:2005rh}. In order for the Q-balls to be formed, it is necessary that the potential for the AD scalar is flatter than $|\psi|^2$ at large field values. After taking account the one-loop correction, the potential for the AD scalar looks like
\begin{equation}
V_{AD,1-loop}(\psi) \sim m^2_{\psi}|\psi|^2 \left( 1+K \ln \frac{|\psi|^2}{M^2} \right)+ \dots,
\end{equation}
where the coefficient $K$ is determined from the renormalization group equations; see, e.g., \cite{Nilles:1983ge, Enqvist:2000gq, Enqvist:1997si}. Loops containing gauginos make a negative contribution proportional to $m^2_{1/2}$, while loops containing sfermions make a positive contribution proportional to $m^2_{AD}$. Thus, when the spectrum is such that the gauginos are much lighter than the sfermions, i.e., $m_{AD} \gg m_{1/2}$, $K$ is likely to be positive and thus Q-balls will not be formed. A more complete analysis of the Q-balls is beyond the scope of the current paper and is left to future investigations.

\subsection{$p=3$ case}
\begin{figure}[t]
  \centering
\includegraphics[width=0.6\textwidth]{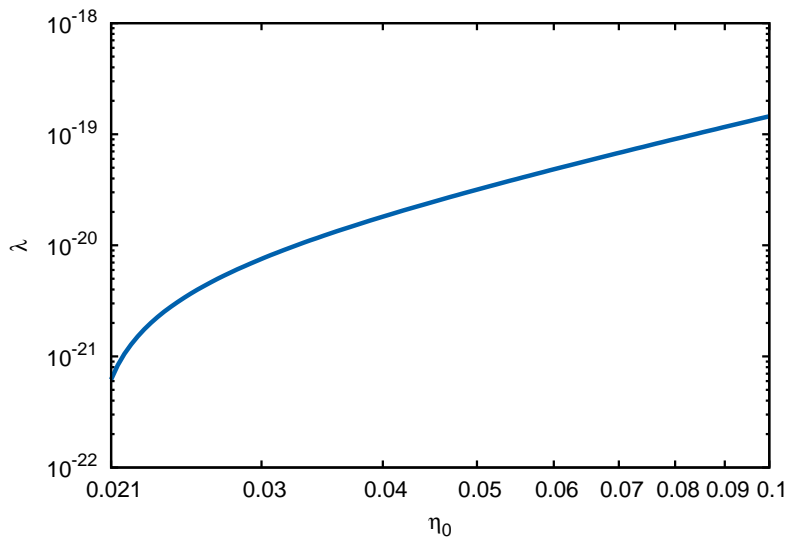}
  \caption{$\lambda$ as a function of $\eta_0$ for $V_0^{1/2}=10^{-12}$ for $p=3$.}
  \label{fig01}
\end{figure}

For $p=3$, we have
\begin{equation}
V_A=\lambda M_P \psi^3.
\end{equation}
From Eq.~(\ref{a6}), by imposing CMB normalization $P_R^{1/2}=5 \times 10^{-5}$ and $n_s=0.96$ we have
\begin{equation}
\lambda=1.51\times 10^{-5}\frac{V_0^{1/2}}{M_P^2}(\eta_0^2-0.0004).
\label{p3lambda}
\end{equation}
For the case $V_0^{1/2}=10^{-12}$ which corresponds to $100$ TeV SUSY breaking scale,
 the parameter $\lambda$ as a function of $\eta_0$ is plotted in Fig.~\ref{fig01}. We can see from the plot that a small value of $\lambda$ and a value of $\eta_0$ which is not far from unity are required. Therefore even if there is a Hubble induced mass, as long as it is not very large, our results are not affected. As we will see, this also applies to $p=4$ and $p=6$ cases.

As an example, if we take $\eta_0=0.03$, we need $\lambda=aA \sim 10^{-20}$.\footnote{It seems we can make $\lambda$ arbitrarily small by fine-tuning $\eta_0$ to approach $\eta_0=0.02$ because when $\eta_0=0.02$, $\lambda=0$ as can be seen from Eq.~(\ref{p3lambda}). However, this is not correct since when $\lambda=0$ the inflaton potential is concave upward and the condition $n_s=0.96$ cannot be achieved. The reason behind this discrepancy is that in this case the small field approximation breaks down and Eq.~(\ref{p3lambda}) is no longer valid.} If we assume
  $A \sim 100\mbox{ TeV} \sim 10^{-13} M_P$, we need the coupling constant $a \sim 10^{-7}$. Interestingly, such small coupling is also required in order to evade rapid proton decay if the $p=3$ flat direction breaks $B-L$ symmetry. For the case $V_0^{1/2}=10^{-14}$ which corresponds to TeV SUSY breaking scale, the corresponding $\lambda$ is $0.01$ times smaller.

From Eq.~(\ref{reheating}), for $p=3$, we have
\begin{equation}
T_R= 1.2 \times 10^{-10} \frac{H^3 M_P}{\lambda \psi_{end}^3}.
\label{eq28}
\end{equation}
From Eqs.~(\ref{a7}) and (\ref{a8}), we have
\begin{equation}
\left( \frac{\psi}{M_P} \right)=\left(\frac{V_0}{M_P^4} \right)\frac{\eta_0+0.02}{6\lambda}
\label{pn60}
\end{equation}
and 
\begin{equation}
\psi_{end}=\frac{2\eta_0}{(\eta_0-0.02)e^{60\eta_0}+\eta_0+0.02}\psi(N=60).
\label{psiend3}
\end{equation}

\subsection{$p=4$ case}
\begin{figure}[t]
  \centering
\includegraphics[width=0.6\textwidth]{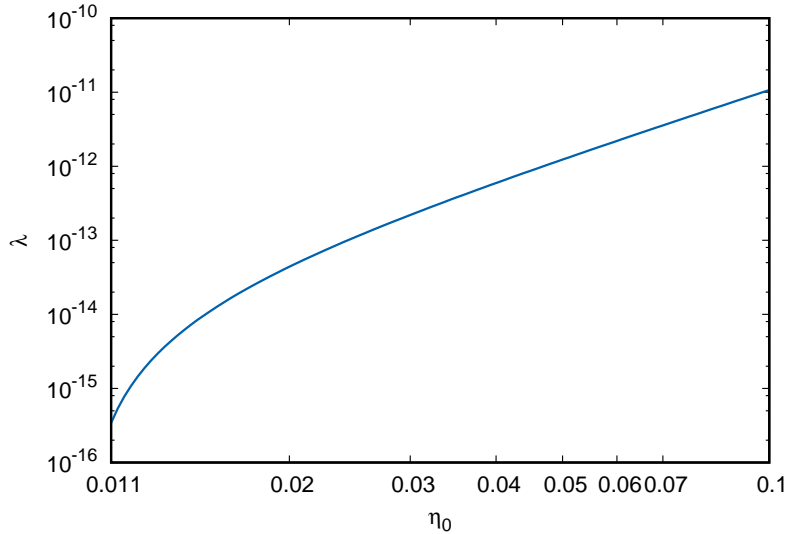}
  \caption{$\lambda$ as a function of $\eta_0$ for $p=4$. Note that here $\lambda$ is independent of $V_0$}
  \label{fig001}
\end{figure}

For $p=4$, we have
\begin{equation}
V_A= \lambda \psi^4.
\end{equation}
From Eq.~(\ref{a6}), by imposing CMB normalization $P_R^{1/2}=5 \times 10^{-5}$ and $n_s=0.96$ we have
\begin{equation}
\lambda=1.10\times 10^{-8}(\eta_0+0.02)(\eta_0-0.01)^2.
\label{p4lambda}
\end{equation}
Note that for $p=4$, $\lambda$ does not depend on $V_0$. The parameter $\lambda$ as a function of $\eta_0$ is plotted in Fig.~\ref{fig001}. We can see from the plot that a small value of $\lambda$ is required. However, since $\lambda$ is the ratio of SUSY breaking A-term and the Planck mass, its value is naturally small.

From Eq.~(\ref{reheating}), for $p=4$, we have
\begin{equation}
T_R= 1.2 \times 10^{-10} \frac{H^3 M_P^2}{\lambda \psi_{end}^4}.
\label{eq281}
\end{equation}
From Eqs.~(\ref{a7}) and (\ref{a8}), we have
\begin{equation}
\left( \frac{\psi}{M_P} \right)^2=\left(\frac{V_0}{M_P^4} \right)\frac{\eta_0+0.02}{12\lambda}
\label{pn60}
\end{equation}
and 
\begin{equation}
\psi_{end}^2=\frac{3\eta_0}{(2\eta_0-0.02)e^{120\eta_0}+\eta_0+0.02}\psi^2(N=60).
\label{psiend4}
\end{equation}

\subsection{$p=6$ case}

\begin{figure}[t]
  \centering
\includegraphics[width=0.6\textwidth]{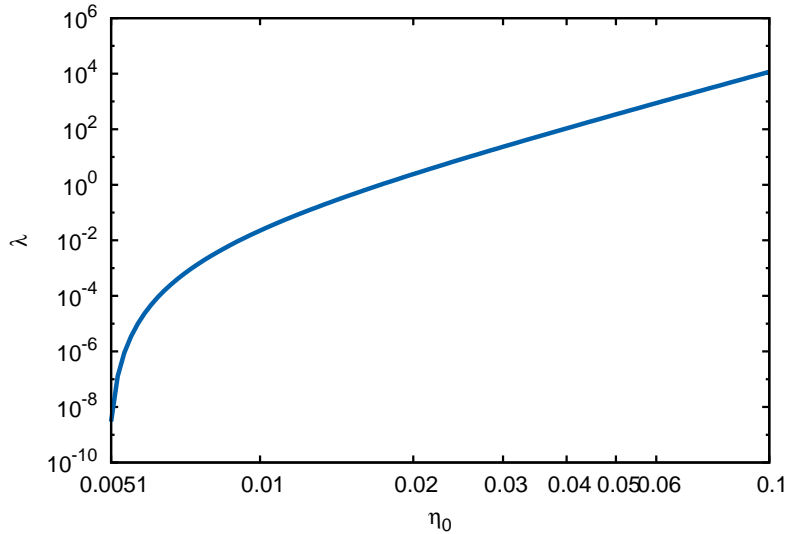}
  \caption{$\lambda$ as a function of $\eta_0$ for $V_0^{1/2}=10^{-12}$ for $p=6$.}
  \label{fig03}
\end{figure}

For $p=6$, we have
\begin{equation}
V_A=\frac{aA}{M_P}\frac{\psi^6}{M_P^2} \equiv \lambda \frac{\psi^6}{M_P^2}.
\label{eq31}
\end{equation}
From Eq.~(\ref{a6}), by imposing CMB normalization $P_R^{1/2}=5 \times 10^{-5}$ and $n_s=0.96$ we have
\begin{equation}
\lambda=7.46\times 10^{-17} \left( \frac{M_P^4}{V_0} \right)  (\eta_0+0.02)(2\eta_0-0.01)^4.
\label{p6lambda}
\end{equation}
For the case $V_0^{1/2}=10^{-12}$ which corresponds to $100$ TeV SUSY breaking scale, the parameter $\lambda$ as a function of $\eta_0$ is plotted in Fig.~\ref{fig03}.
From the plot, we can see that $\lambda$ is not necessarily a small number and can be quite large depending on $\eta_0$. It may be interesting to note that $\lambda \sim \mathcal{O}(1)$ can be achieved. On the other hand, from Eq.~(\ref{eq31}), we can see that it seems for TeV $<A<100$ TeV, $\lambda$ has to be very small unless coupling $a$ is very large\footnote{A larger effective $\lambda$ may be obtained if we introduce a singlet $\chi$ with the superpotential $W_{\psi,\chi}=\frac{a}{M_P}\psi^3\chi+\frac{1}{2}M_\chi \chi^2$. We will leave the detailed model building in our future work.}.  
From Eq.~(\ref{reheating}), for $p=6$, we have
\begin{equation}
T_R=1.2 \times 10^{-10}\frac{H^3M_P^4}{ \lambda \psi_{end}^6}.
\label{eq33}
\end{equation}
From Eqs.~(\ref{a7}) and (\ref{a8}), we have
\begin{equation}
\left( \frac{\psi}{M_P} \right)^4=\left(\frac{V_0}{M_P^4} \right)\frac{\eta_0+0.02}{30\lambda}
\label{pn60}
\end{equation}
and 
\begin{equation}
\psi_{end}^4=\frac{5\eta_0}{(4\eta_0-0.02)e^{240\eta_0}+\eta_0+0.02}\psi^4(N=60).
\label{psiend6}
\end{equation}

\subsection{Discussion}
\begin{figure}[t]
  \centering
\includegraphics[width=0.6\textwidth]{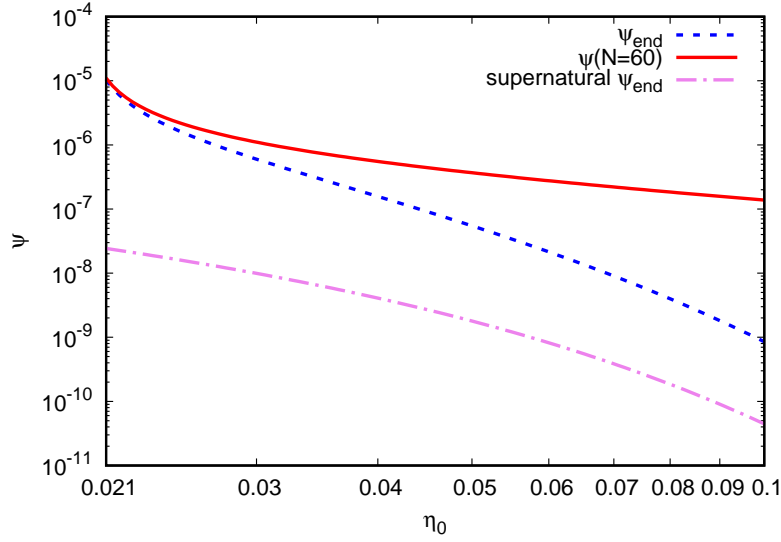}
  \caption{$\psi/M_P$ as a function of $\eta_0$ for $V_0^{1/2}=10^{-12}$ and $p=3$. The field value at the end of supernatural inflation is plotted for comparison.}
  \label{fig02}
\end{figure}

\begin{figure}[t]
  \centering
\includegraphics[width=0.6\textwidth]{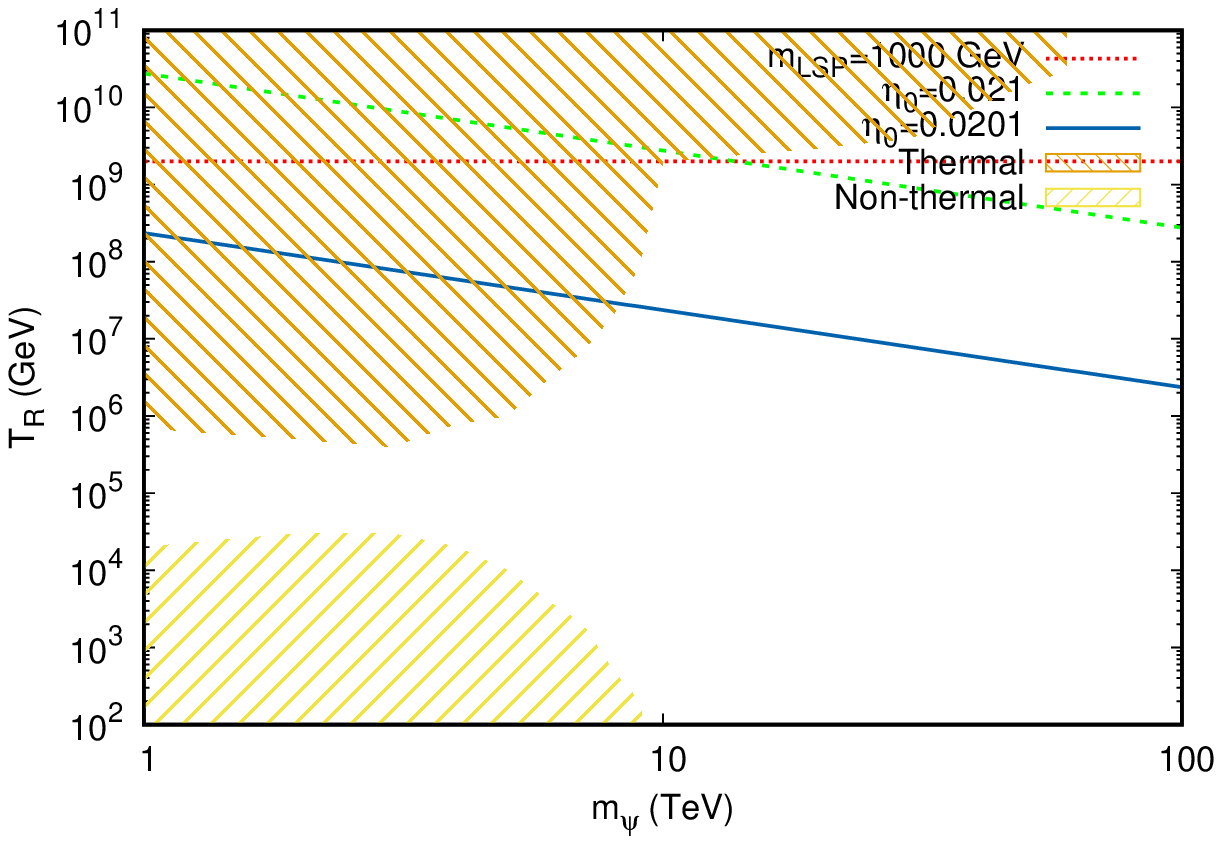}
  \caption{$T_R$ as a function of $m_\psi$ for $\eta_0=0.03$ and $\eta_0=0.021$ in the case of $p=3$. Here it is assumed that $m_\psi = m_{3/2}$. If $m_\psi < m_{3/2}$, the gravitino bound shifts to the left and becomes weaker or even negligible. We also include a bound of LSP production from Eq.~(\ref{lsp}). }
  \label{reheating1}
\end{figure}

\begin{figure}[t]
  \centering
\includegraphics[width=0.6\textwidth]{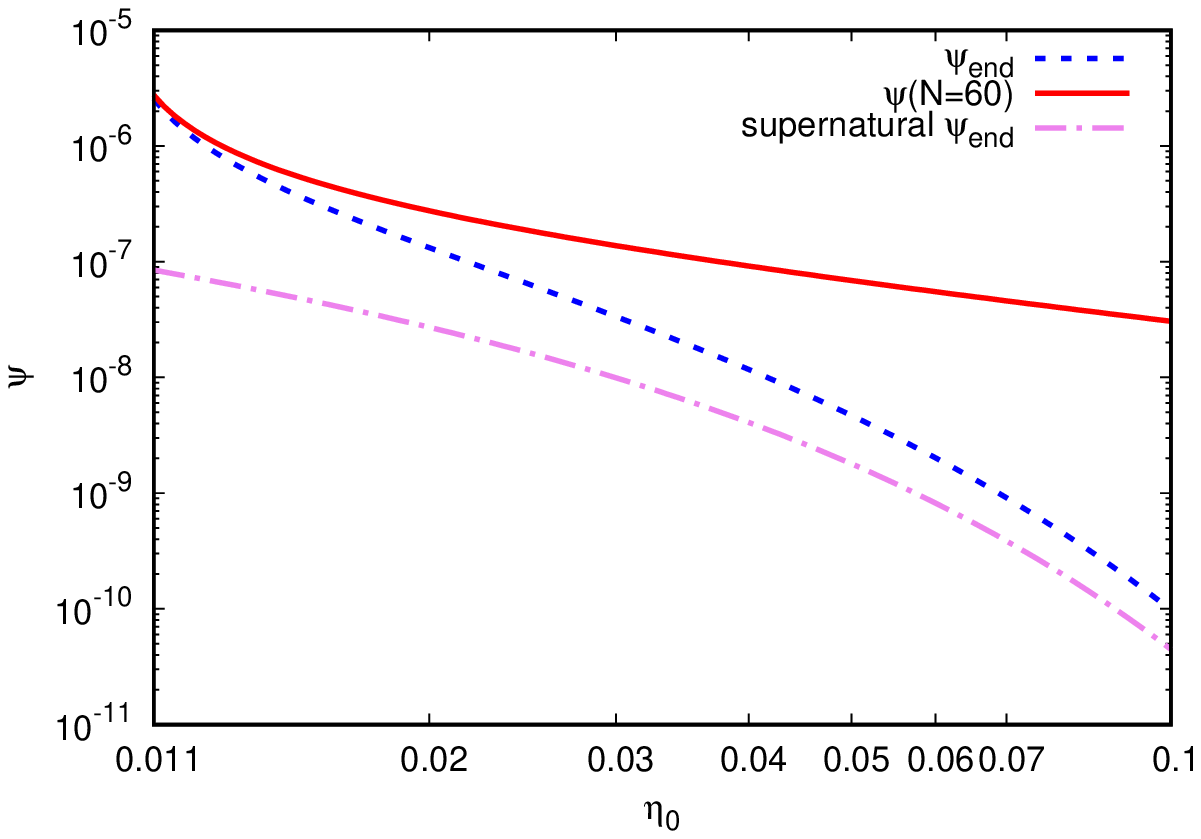}
  \caption{$\psi/M_P$ as a function of $\eta_0$ for $V_0^{1/2}=10^{-12}$ and $p=4$. The field value at the end of supernatural inflation is plotted for comparison.}
  \label{fig021}
\end{figure}

\begin{figure}[t]
  \centering
\includegraphics[width=0.6\textwidth]{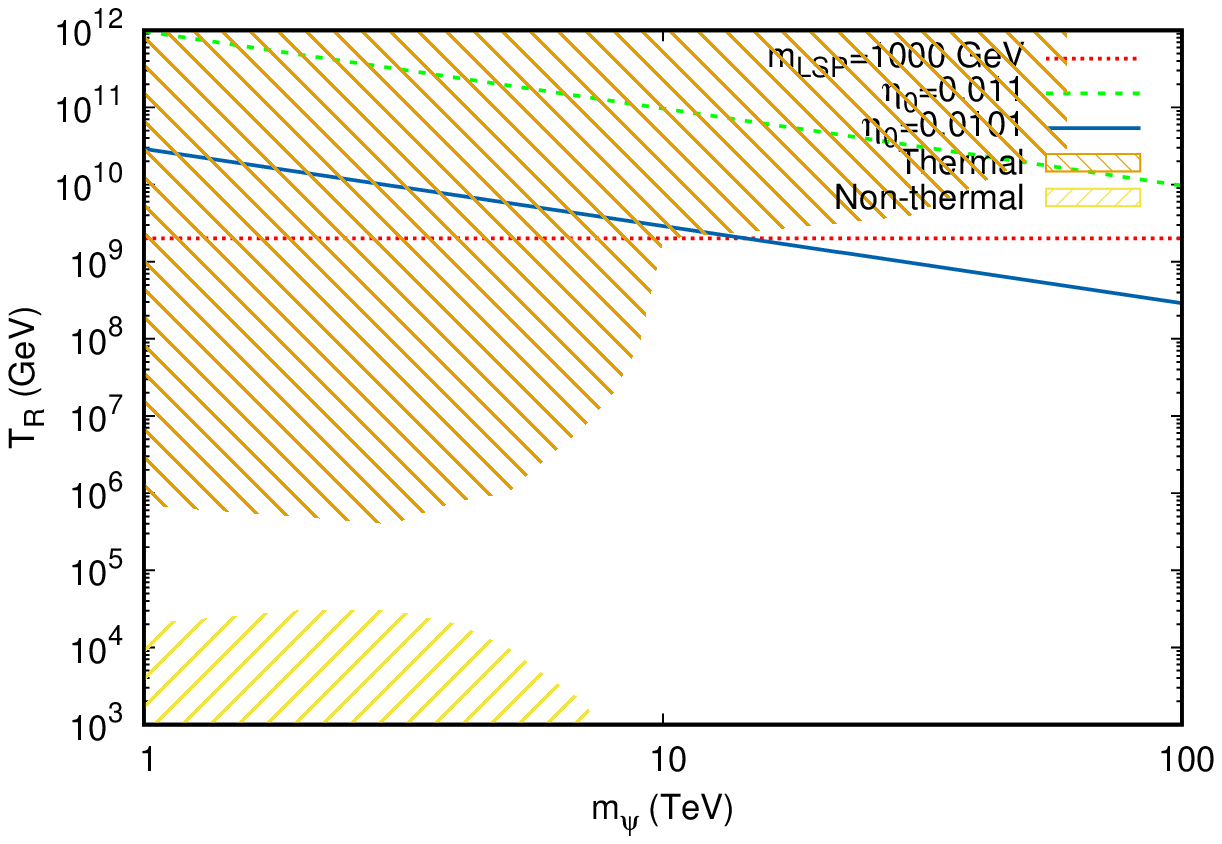}
  \caption{$T_R$ as a function of $m_\psi$ for $\eta_0=0.02$ and $\eta_0=0.011$ in the case of $p=4$. Here it is assumed that $m_\psi = m_{3/2}$. If $m_\psi < m_{3/2}$, the gravitino bound shifts to the left and becomes weaker or even negligible. We also include a bound of LSP production from Eq.~(\ref{lsp}).}
  \label{reheating11}
\end{figure}

\begin{figure}[t]
  \centering
\includegraphics[width=0.6\textwidth]{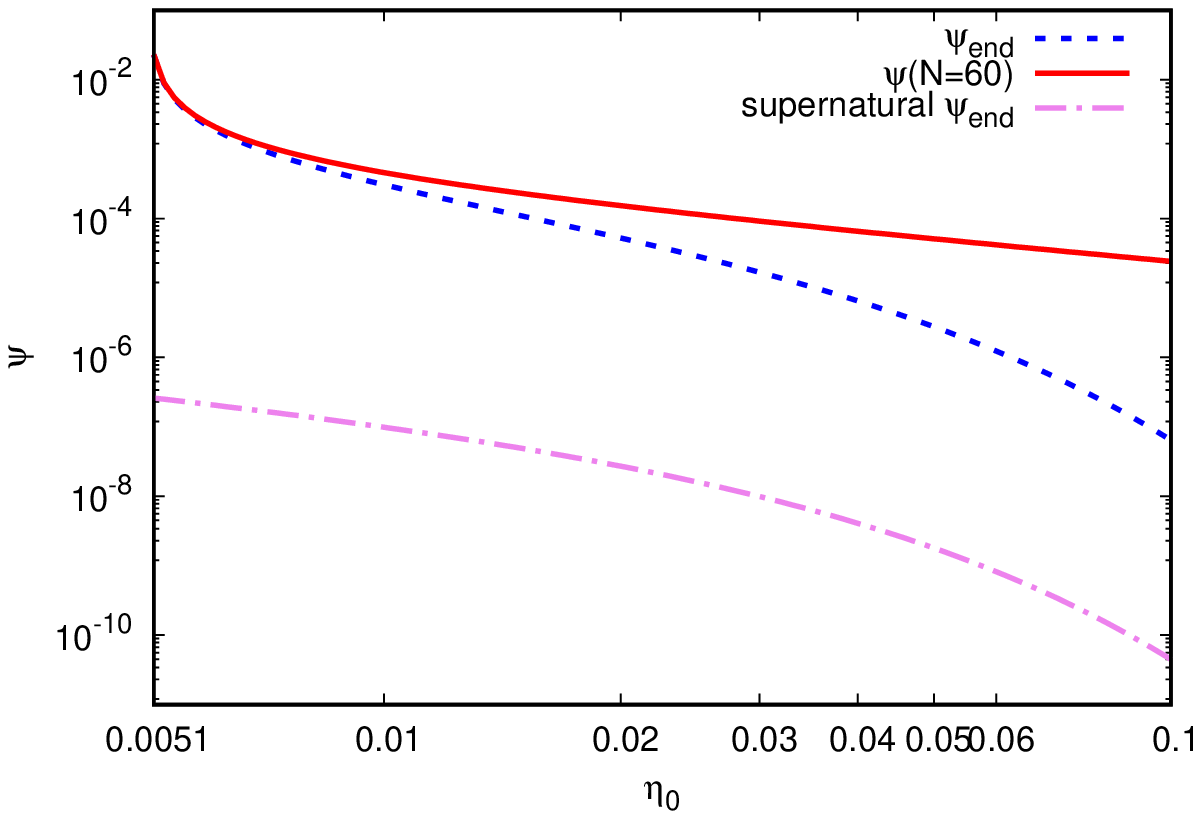}
  \caption{$\psi/M_P$ as a function of $\eta_0$ for $V_0^{1/2}=10^{-12}$ and $p=6$. The field value at the end of supernatural inflation is plotted for comparison.}
  \label{fig04}
\end{figure}

\begin{figure}[t]
  \centering
\includegraphics[width=0.6\textwidth]{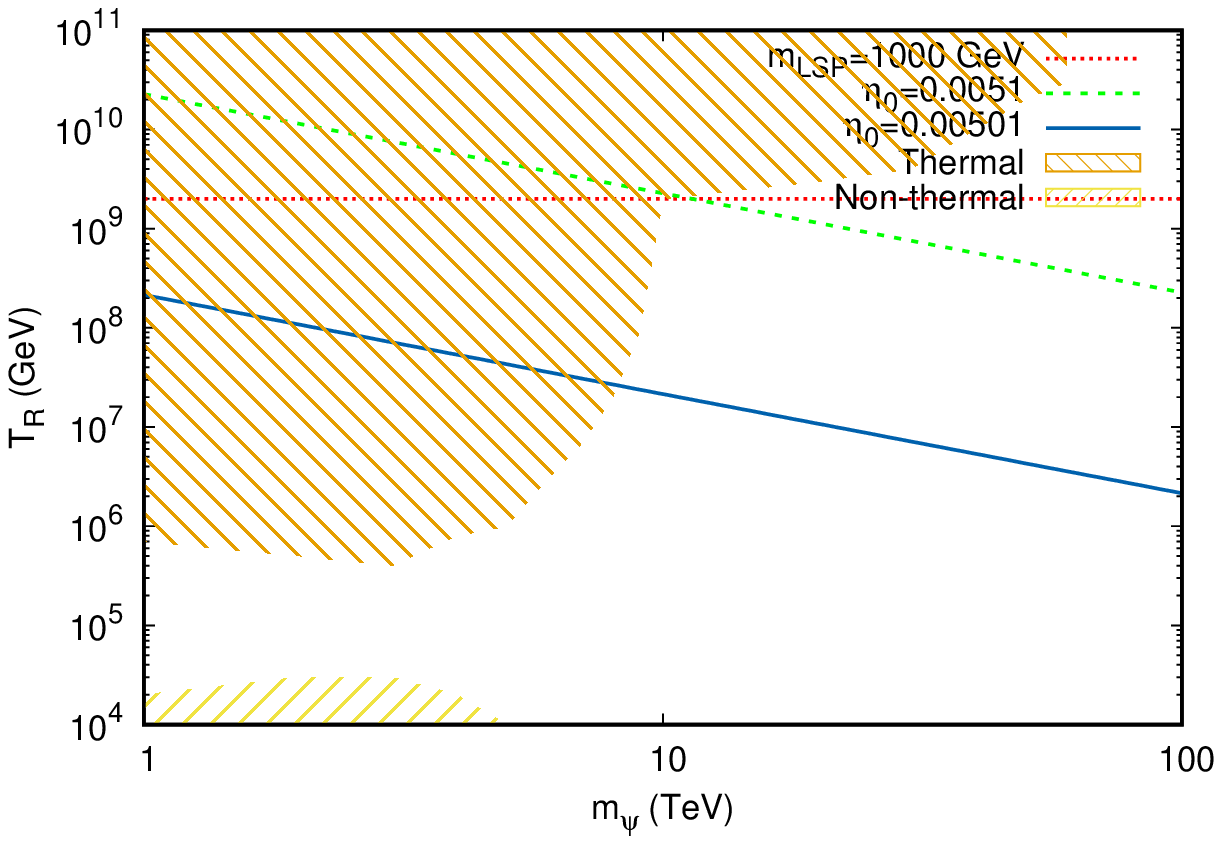}
  \caption{$T_R$ as a function of $m_\psi$ for $\eta_0=0.01$ and $\eta_0=0.0051$ in the case of $p=6$. Here it is assumed that $m_\psi = m_{3/2}$. If $m_\psi < m_{3/2}$, the gravitino bound shifts to the left and becomes weaker or even negligible. We also include a bound of LSP production from Eq.~(\ref{lsp}).}
  \label{reheating2}
\end{figure}

For the case $V_0^{1/2}=10^{-12}$ which corresponds to $100$ TeV SUSY breaking scale, the field values $\psi/M_P$ both at $N=60$ and at the end of inflation as a function of $\eta_0$ are plotted in Figs.~\ref{fig02}, \ref{fig021}, and \ref{fig04}. We also plot the field value at the end of supernatural inflation by using Eqs.~(\ref{sne}) and (\ref{snp}) for comparison. As can be seen from the plot, $\psi_{end}$ of supernatural inflation is smaller than $\psi_{end}$ of hilltop supernatural inflation. Since the produced baryon number is proportional to $\psi_{end}^p$ from Eq.~(\ref{baryon}), this is the reason why AD baryogenesis can work better in the framework of hilltop supernatural inflation.
By using Eqs.~(\ref{eq28}), (\ref{psiend3}), (\ref{eq281}), (\ref{psiend4}), (\ref{eq33}), (\ref{psiend6}), the relations $V_0^{1/2}=m_\psi M_P/\sqrt{\eta_0}$, and $H=V_0^{1/2}/(\sqrt{3}M_P)$, we can obtain the required reheating temperature for successful AD baryogenesis as a function of the inflaton mass $m_\psi$. The results are shown in Figs.~\ref{reheating1}, \ref{reheating11}, \ref{reheating2}. We also include upper and lower bounds for the reheating temperature that correspond, respectively, to thermal and nonthermal gravitino production \cite{Kohri:2010sj, Kawasaki:2006gs, Dine:2006ii, Kawasaki:2006hm, Endo:2006qk, Kawasaki:2006mb, Endo:2007ih, Takahashi:2007gw}. Here it is assumed that $m_\psi=m_{3/2}$. In SUSY breaking scenarios where the gravitino mass is much larger than squark / slepton masses, the plots of the constraints would shift to the left and tend to become weaker. Therefore the constraint we use here is conservative. As we can see from the figure, generally speaking, for higher soft mass and SUSY breaking scale, the required reheating temperature is lower. Note that if we choose a larger $\eta_0$ which corresponds to a smaller $\psi_{end}/M_P$, we will need a larger reheating temperature. We can also note from the figure, the reheating temperature predicted for $p=4$ appears to be higher than that for $p=3$ or $p=6$.

For $p=3$, if we choose $\eta_0$ approaches $0.02$, $\psi_{end}/M_P$ can be bigger and a lower reheating temperature results. However, as $\eta_0$ approaches $0.02$, $\psi_{end}/M_P$ becomes sensitive to $\eta_0$. As can be seen from Eqs.~(\ref{p3lambda}) and (\ref{pn60}), $\eta_0=0.02$ cannot be achieved because the formula becomes singular and our small field approximation breaks down. Similar behavior occurs for $p=4$ and $p=6$ when $\eta_0$ approaches $0.01$ and $0.005$ respectively.

\section{Baryon isocurvature perturbation}

Since the AD field (as the inflaton in our model) is a complex field, it is possible that during inflation the quantum fluctuations of the phase would induce isocurvature perturbation. The fluctuations of the phase of the AD field are given by
\begin{equation}
\delta \theta = \frac{H}{2\pi \psi},
\end{equation}  
where $H$ and $\psi$ are the values obtained during inflation at $N=60$. The baryon isocurvature perturbation is defined as
\begin{equation}
S_{b\gamma} \equiv \frac{\delta\rho_B}{\rho_B}-\frac{3}{4}\frac{\delta \rho_\gamma}{\rho_\gamma}=\delta \log \left(\frac{\rho_B}{s}\right),
\end{equation}
where $\rho_B$ and $\rho_\gamma$ are the energy densities of the baryons and photons. From Eq.~(\ref{baryon}), we have
\begin{equation}
 S_{b\gamma}=p\cot[p\theta_{sp}+arg(A)]\frac{H}{2\pi\psi}\sim p\frac{H}{2\pi\psi}.
\label{iso}
\end{equation}
In the second equality of the above equation, we assume $\cot[p\theta_{sp}+arg(A)] \sim \mathcal{O}(1)$ . From the latest Planck 2018 data for dark matter isocurvature perturbation $S_{c\gamma}$     
 \cite{Akrami:2018odb},
\begin{equation}
S_{c\gamma}=\frac{\Omega_b}{\Omega_c}S_{b\gamma} \lesssim \left( \frac{\beta_{iso}}{1-\beta_{iso}}P_R\right)^{\frac{1}{2}}=10^{-5},
\end{equation}
where $\beta_{iso}<0.038$ is used. We can obtain
\begin{equation}
S_{b\gamma} \lesssim 5.33\times 10^{-5},
\end{equation}
where we have used $\Omega_c/\Omega_b=5.33$. By using Eq.~(\ref{iso}) and $V_0^{\frac{1}{2}}=\sqrt{3}HM_P=10^{-12}$, we obtain
\begin{equation}
\psi(N=60) \gtrsim 1.7 \times 10^{-9}p.
\label{iso2}
\end{equation}
From Figs.~\ref{fig02}, \ref{fig021}, and \ref{fig04} we can see that in our parameter space, Eq.~(\ref{iso2}) is satisfied and there is no observable baryon isocurvature perturbation produced.

\section{Dark matter production}

If the reheating temperature required for successful AD baryogenesis is high, and the parameter space is such that thermal gravitino problem is evaded, e.g., large gravitino mass, there is another upper bound for the reheating temperature from LSP production given by 
\begin{equation}
T_R<2\times 10^{10}\mbox{ GeV} \left( \frac{100\mbox{ GeV}}{m_{LSP}} \right).
\label{lsp}
\end{equation} 
It suggests an interesting possibility that when the reheating temperature required for successful baryogenesis is high, dark matter can also be generated.

There is another possible way to generate dark matter in our model.
For higher scale SUSY breaking, the required reheating temperature is lower; this helps to evade gravitino problem and if we choose a $100$ TeV inflaton mass, the decay could lead to a nonthermal origin for dark matter \cite{Moroi:1999zb,Acharya:2008bk,Acharya:2009zt, Acharya:2010af, Kane:2011ih, Kim:2016spf}. In addition to the nonthermal production of the LSP from the decaying
gravitino, there should be a component of the LSP produced in the thermal plasma. If $T_R > m_{LSP} / 25$,  there is a component to be the standard thermal relic. If $T_R < m_{LSP} / 25$, we need a nonthermal annihilation with the cross section which is much larger than the canonical one ($ \langle\sigma v\rangle \sim  3 \times 10^{-26}  \mbox{cm}^3  $/ sec).




A candidate of CDM can be Wino (or gaugino). Even for the freeze-out scenario, the canonical annihilation cross section ($ \langle\sigma v\rangle \sim  3 \times 10^{-26}  \mbox{cm}^3  $/ sec) is realized only at one point ($m_{LSP} = 3$ TeV).  Except for that point, the thermal relic for wino cannot explain the 100\% of CDM.

For $m_{LSP} < 3$ TeV, we need an additional component by nonthermal production of Wino by a decaying long-lived particle such as gravitino (this scenario) because the thermal relic through the freeze-out is short to the observed $\Omega_{CDM}$.

For $m_{LSP} > 3$ TeV, we need entropy production to dilute the thermal relic of wino by sizable amount of decaying gravitino or modulus and so on.

\section{Conclusion}
\label{con}
In this paper, we have shown that it is possible for the inflaton field to play the role of the Affleck-Dine field to produce successful baryogenesis in the model of hilltop supernatural inflation. We have considered the cases realized via SUSY flat directions $p=3$, $p=4$, $p=6$. Since hilltop supernatural inflation belongs to the category of small field inflation, the tensor-to-scalar ratio would not be observable in near future experiments. We have to find further experimental results to distinguish which case would be better, perhaps from particle physics since these different cases can be connected to different particle physics phenomena. Depending on the parameters and the resulting reheating temperature, both the thermal and nonthermal gravitino bound can be satisfied. We calculated the baryon isocurvature perturbation and found that it can be neglected for all the cases throughout the parameter space. We also explore the interesting possibility that the reheating temperature for successful baryogenesis can also be responsible for LSP dark matter production. If the inflaton mass is around $100$ TeV, dark matter could have been produced nonthermally via inflaton decay. Generally speaking, the scenario of dark matter production depends on the mass of the candidate of CDM. Comparing with other inflation models, we have shown that hilltop supernature inflation has a rich connection to particle physics which can be further explored in the future research.

\appendix
\section{analysis of hilltop supernatural inflation}

For hilltop supernatural inflation with the potential given by Eq.~(\ref{Vpsi}), the field value during inflation is 
\begin{eqnarray}
\left(\frac{\psi}{M_P}\right)^{p-2}&=&\left(\frac{V_0}{M_P^4}\right) 
\frac{\eta_0 e^{(p-2)N\eta_0}}{\eta_0 x+p \lambda (e^{(p-2)N\eta_0}-1)} \label{a2}\\
x &
\equiv & \left(\frac{V_0}{M_P^4}\right) \left(\frac{M_P}{\psi_{end}}\right)^{p-2}. \label{a3}
\end{eqnarray}
The spectrum and the spectral index are
\begin{eqnarray}
P_R&=&\frac{1}{12\pi^2} \left(\frac{V_0}{M_P^4}\right)^{\frac{p-4}{p-2}} e^{-2N\eta_0}\frac{[p\lambda(e^{(p-2)N\eta_0}-1)+\eta_0 x]^{\frac{2p-2}{p-2}}}{\eta_0^{\frac{2p-2}{p-2}}(\eta_0 x-p\lambda)^2} \label{a4},\\
n_s&=&1+2\eta_0 \left[1-\frac{\lambda p(p-1) e^{(p-2)N\eta_0}}{\eta_0 x+p\lambda(e^{(p-2)N\eta_0}-1)}\right]. \label{a5}
\end{eqnarray}
We can compare Eq.~(\ref{a5}) with Eq.~(\ref{snn}).
From the above equations, we can obtain 
\begin{equation}
\lambda=\frac{(12 \pi^2 P_R)^{\frac{p-2}{2}}}{p[2(p-1)]^{(p-1)}} \left( \frac{V_0}{M_P^4} \right)^{-\frac{p-4}{2}}(2\eta_0-n_s+1)(2(p-2)\eta_0+n_s-1)^{(p-2)}.
\label{a6}
\end{equation}
The field value during inflation and at the end of inflation can be obtained from Eqs.~(\ref{a2}), (\ref{a3}) and (\ref{a5}) as
\begin{equation}
\left( \frac{\psi}{M_P} \right)^{p-2} = \left( \frac{V_0}{M_P^4} \right) \frac{\eta_0+\frac{1-n_s}{2}}{\lambda p(p-1)}
\label{a7}
\end{equation}
and 
\begin{equation}
\left( \frac{\psi}{\psi_{end}} \right)^{p-2}=\frac{[2\eta_0p-4\eta_0+(n_s-1)]e^{(p-2)N\eta_0}+2\eta_0-n_s+1}{2\eta_0(p-1)}.
\label{a8}
\end{equation}

\acknowledgments
This work was supported in part by JSPS KAKENHI Grant No.~JP17H01131 (K.K.), MEXT Grant-in-Aid for Scientific Research on Innovative Areas Grants No. JP15H05889 (K.K.), No.  JP18H04594 (K.K.), and No. JP19H05114 (K.K.), the Ministry of Science and Technology (MOST) of Taiwan under Grant No. MOST 109-2112-M-167-001 (C.-M.L).
We would like to thank Yukihiro Mimura for useful discussions and comments.

\end{document}